\documentclass[12pt,preprint]{aastex}
\shortauthors{Ramanpreet Kaur et al.}
\shorttitle{QPOs in XTE J0111.2--7317}

\slugcomment{Submitted for publication in The Astrophysical Journal}
\received{}

\begin{document}

\title{Quasi periodic oscillations in XTE J0111.2--7317, highest frequency
among the HMXB pulsars}

\author{Ramanpreet Kaur\altaffilmark{1},
 Biswajit Paul\altaffilmark{2}, Harsha Raichur\altaffilmark{2} and Ram Sagar\altaffilmark{1}}

\altaffiltext{1}{Aryabhatta Research Institute of Observational Sciences, Naini Tal
263\,129, India}
\altaffiltext{2}{Raman Research Institute, Bangalore 560\,080, India}

\begin{abstract}
We report here discovery of Quasi Periodic Oscillations (QPOs) in the
High Mass X-ray Binary (HMXB) Pulsar XTE J0111.20--7317 during a
transient outburst in this source in December 1998. Using observations
made with the proportional counter array of the Rossi X-ray Timing
Explorer during the second peak and the declining phase of this
outburst we have discovered a QPO feature at a frequency of 1.27 Hz.
We have ruled out the possibility that the observed QPOs can instead be 
from the neighbouring bright X-ray pulsar SMC X-1.
This is the highest frequency
QPO feature ever detected in any HMXB pulsar. In the absence of a
cyclotron absorption feature in the X-ray spectrum, the QPO feature,
along with the pulse period and X-ray flux measurement measurement
helps us to constrain the magnetic field strength of the neutron star.
\end{abstract}

\section{Introduction}

Quasi periodic oscillations (QPOs) in X-ray binary pulsars are thought to be
related to the motion of inhomogeneous matter distribution in the inner
accretion disk and give us useful information about the interaction between
accretion disks and the central object. The frequencies of these oscillations
in X-ray pulsars (excluding the millisecond accreting pulsars) range from
$\simeq$ 1 mHz to $\simeq$ 40 Hz and they can be from $\simeq$ 100 times
smaller to $\simeq$ 100 times larger than the pulsar spin frequencies
(Psaltis D. 2004). Owing to different behavior of the QPO features in different
sources, especially its relation with the X-ray luminosity, it is not yet
certain whether the QPOs in all the X-ray pulsars arise due to the same
mechanism. Investigation of the QPOs and its variations with photon
energy and luminosity state, therefore, gives us important clues about the
extent and structure of the disk and QPO generation mechanism in accreting X-ray
pulsars.

The transient X-ray pulsar XTE J0111.2--7317 was discovered with the
Proportional Counter Array (PCA) of the Rossi X-ray Timing Explorer
(RXTE) in November 1998 (Chakrabarty et al. 1998a) and was 
simultaneously detected in hard X-rays with the Burst and Transient
Source Experiment (BATSE) on board the Compton Gamma Ray Observatory
(CGRO) with a flux ranging from 18 to 37 mCrab (Wilson \& Finger 1998).
Public data of CGRO/BATSE and RXTE/ASM revealed that this source was
seen in outburst in both hard and soft X-ray during 1998 November-1999
January. Follow-up observations were taken up by Advanced Satellite for 
Cosmology and Astrophysics (ASCA) to study the pulsations and the X-ray 
spectrum extending upto 10 keV and thus detected it with a flux of 
3.6$\times$10$^{-10}$ erg s$^{-1}$cm$^{-2}$ in the 0.7 - 10.0 keV band 
(Chakrabarty et al. 1998b; Yokogawa et al. 2000). The ASCA observations
also revealed the presence of a pulsating soft excess, which subsequently
led to detailed investigations of a similar feature in several accreting
X-ray pulsars Her X-1, LMC X-4, SMC X-1 etc. (Endo, Nagase \& Mihara 2000, 
Paul et al. 2002, Naik \& Paul 2004a,b) and is now understood to be due to
reprocessing of the hard X-rays from the inner accretion disk (Hickox,
Narayan \& Kallman 2004). ASCA observations during the outburst also gave
the opportunity to find 
position of the source with an error circle of 15$\arcsec$. BATSE
observations found the pulsar to be spinning up with a short 
time-scale of $\sim$20 years (Yokogawa et al. 2000), which confirms that 
the compact object is a neutron star. This object is present in the
direction of the Small Magellanic Cloud (SMC), and it is very likely
that it belongs to SMC (Yokogawa et al. 2000). The SMC association
was later on confirmed (Coe et al. 1998) by finding the average
velocity shift of optical lines to be 166 $\pm$ 15 km s$^{-1}$ which is
comparable with $\sim$ 166 km s$^{-1}$ for SMC (Feast et al. 1961).
Optical counterpart of XTE J0111.2--7317 was first proposed to be a B
star with strong H$\alpha$ and H$\beta$ emission (Israel et al. 1999)
and later on confirmed to be a B0.5-B1Ve star (Covino et al. 2001).

In the subsequent sections we describe timing analysis of the archival
X-ray data of XTE J0111.2--7317 from RXTE and we report the discovery of
a QPO feature from this source. We investigated the possibility of the
QPOs arising
from the nearby bright X-ray pulsar SMC X-1 and discuss the implications
of the QPOs in this source, especially regarding the strength of the
neutron star magnetic field.
 
\section{Observations and Analysis}

XTE J0111.2--7317 went into an outburst in November 1998 and was discovered
with RXTE-PCA during scans of the SMC X-1 region (Chakrabarty et al. 1998a).
Subsequently, two short observations
of the source were carried out on 18 December with RXTE-PCA and later
the source was monitored frequently from 22 December 1998 to 19 February
1999 as a target of opportunity. There were twenty pointed observations
during this time, each with an exposure of $\sim$ 2-3 ks. For most of the 
pointings all the five proportional counter units (PCUs) were ON whereas
on some occasions three to four PCUs were ON. The source was also regularly 
monitored by All Sky Monitor (ASM) on board RXTE. The nearby bright binary
X-ray pulsar SMC X-1 is only 30$\arcmin$ away from XTE J0111.2--7317 and it
falls in the field of view (FOV) of RXTE-PCA during observations of
XTE J0111.2--7317. Thus we have also used the RXTE-ASM data of SMC X-1
available for the outburst period of XTE J0111.2--7317 to know its flux
contributions to the XTE J0111.2--7317 lightcurve and the power density spectrum.
Long term lightcurves of XTE J0111.2--7317 and SMC X-1 measured with the
RXTE-ASM are shown in Figure 1 for $\sim$300 days, covering the outburst
of XTE J0111.2--7317 and about five super-orbital intensity modulations of SMC X-1.
Lightcurve of XTE J0111.2-7317 (contaminated by SMC X-1 in parts) taken
with RXTE-PCA is also shown in the same Figure with a different marker and
different normalization. 

Lightcurves were extracted from observations of XTE J0111.2--7317 with the
RXTE-PCA with a time resolution of 0.125 s using the Standard-1 data. The
background count rates were simulated and subtracted from the Standard-1
lightcurves. The 31 s pulsations of XTE J0111.2--7317 were clearly seen in
lightcurves, except during the last few days. The lightcurves were divided
into small segments each of length 1024 s and a power density spectrum of
each segment was generated. The power spectra were normalised such that
their integral gives the squared rms fractional variability and the expected
white noise level was subtracted. Figure 2 shows two power spectra
averaged over the time ranges 'A' and 'B' marked in Figure 1 when at least
one of
the two sources XTE J0111.2--7317 and SMC X-1 was bright. The peak
at $\sim$0.032 Hz and its harmonics seen in the top spectrum of Figure 2
are due to the pulsations of XTE J0111.2--7317. A small hump seen in the
same power spectrum at $\sim$1.27 Hz is a Quasi periodic oscillations feature. 
The spectrum shown in the bottom of Figure 2 (representing the
time range 'B' of Figure 1) shows the absence of 31 s pulsations while the
0.7 s pulsations of SMC X-1 and its harmonics are clearly detected. Figure
3 gives an expanded view of the power spectrum of XTE J0111.2--7317, in the
frequency range of 0.5 to 4.0 Hz. The solid line is the best fitted model
with one component for the continuum and a second Gaussian component for the
QPO feature. From the individual power spectra we found that the QPO
signature was prominent during the time of outburst from 51165 MJD to
51173 MJD and faded as the outburst decayed. Inclusion of the Gaussian
QPO feature in the model reduced the {$\chi^2$} by 77 for 98 degrees of
freedom. The QPO feature is detected with a signal to noise ratio of more
than 9. The average QPO frequency was
measured to be $1.266 \pm 0.018$ Hz with an rms fraction of $2.52 \pm 0.15\%$.
The width of the QPO feature in the Gaussian model was measured to be $\sigma 
=  0.07 \pm 0.01$ Hz, making it one of the narrowest QPO features among 
accretion powered X-ray pulsars. 

Since SMC X-1 lies within the FOV of RXTE-PCA during the XTE J0111.2--7317
observations, the QPO that is seen in the power spectrum of XTE J0111.2--7317
lightcurves could also be a contribution of SMC X-1. Figure 1 shows the
RXTE-ASM lightcurves of SMC X-1 (filled circles) and XTE J0111.2--7317
(open circles) along with the scaled down PCA lightcurve of XTE J0111.2--7317
(crosses) during the 1999 outburst. The SMC X-1 ASM lightcurve clearly shows
the semi-periodic intensity variations of about 60 days, which is supposedly 
its super-orbital period. As can be seen in Figure 1, during the outburst
of XTE J0111.2--7317, the ASM count rate of SMC X-1 was considerably high.
However as shown in Figure 2, the RXTE-PCA lightcurve of XTE J0111.2--7317 shows
the 0.71 s pulsations due to SMC X-1 contamination during the later part
of the observations when the flux of  XTE J0111.2--7317 has decayed (segment
'B' if Figure 1). But during the peak of the outburst of XTE J0111.2--7317, 
from MJD 51165 to 51173 (segment 'A' in Figure 1), the lightcurve obtained 
by PCA does not show the 0.71 s pulsations due to SMC X-1. It is also during 
this time that the QPO feature is most prominent in the power spectra. We 
have invesigated the binary phase of SMC X-1 during the observations of 
XTE J0111.2--7317 reported here and found out that the observations made 
in segment A are during the eclipse of SMC X-1. However the segment B
observations were done when SMC X-1 was in eclipse egress. Figure 4 shows
RXTE-PCA lightcurve of XTE J0111.2--7317 during segment A of Figure 1,
along with the 10 year long RXTE-ASM light curve of SMC X-1, both folded
with the orbital period of SMC X-1 (Paul, Raichur \& Mukherje 2005) during
the same time interval. 

We have also used additional PCA data to estimate the possible level of
contribution
from SMC X-1 by its pulsed X-rays. SMC X-1 was observed extensively by RXTE 
from 24 November 1996 to 5 September 1998 over a range of intensity level of 
the source. We used the event mode data of RXTE-PCA to obtain the lightcurve 
of SMC X-1 during this time with a time resolution of 25 ms. We first measured 
the local spin periods of the pulsar from barycenter corrected lightcurves and 
then created the pulse profiles by folding the lightcurves at the respective 
spin period. The individual
observations had short time spans of less than 3 ks and therefore the smearing
of the pulse profile due to orbital motion of the pulsar was negligible.
The difference between the maximum and minimum count rates in the pulse
profile was taken as a measure of the pulsed X-ray intensity for each
observation. The average X-ray intensity was measured by fitting a constant
to the folded lightcurve. A plot of the pulsed X-ray intensity versus the
average X-ray intensity of SMC X-1 measured from the RXTE-PCA observations
is shown in Figure 5. It can be clearly seen that pulsed X-ray intensity
and the average X-ray intensity are very closely correlated, with a formal
correlation coefficient of 0.97. Below an average
source plus background count rate of $42$ cnt s$^{-1}$ detector$^{-1}$,
pulsations are not detected in SMC X-1. As we did not detect the SMC X-1
pulsations along with the 1.27 HZ QPOs (segment 'A' of Figure 1), we can
separeately conclude that contributions of SMC X-1 towards the total flux
is negligible in segment 'A', and therefore, the QPOs must be a feature of
XTE J0111.2-7317.

\section{Discussion}

We have discovered QPOs from observations of the High Mass X-ray
Binary (HMXB) pulsar XTE J0111.2--7317 during the second peak and declining
phase of its transient outburst in 1998-1999. The two peaks during the 
outburst can be clealy seen in the lightcurve taken by BATSE onboard CGRO
during MJD 51120 to 51200 (Yokogawa et al. 2000). However RXTE-PCA observations 
for XTE J0111.2--7317 were made during the second peak of the outburst 
from MJD 51165 to 51228. The 700 ms pulsations of SMC X-1 are detected
in part of the light curve near the end of the outburst (segment 'B' of
Figure 1) and the corresponding power spectrum is shown in Figure 2.
We have found out that the observations in segment 'B' were made when
SMC X-1 was in eclipse egress. 
We have ruled out the possibility that the QPOs observed during the segment
'A' are from SMC X-1 which would have been equally interesting. QPOs have
been detected in about one dozen accretion powered X-ray pulsars, including
three pulsars with low mass companions, GRO 1744--28 (Zhang et al. 1996),
4U 1626--67 (Shinoda et al. 1990; Kommers et al. 1998) and Her X-1 (Boroson
et al. 2000; Moon \& Eikenberry 2001b; Makishima et al. 1999). Among the HMXB
pulsars, QPOs seem to occur equally frequently in transient and persistent
sources. Transient HMXB pulsars from which QPOs have been detected are
EXO 2030+375 (Angelini et al. 1989), A 0535+262 (Finger et al. 1996),
XTE J1858+034 (B. Paul et al. 1998, Mukherjee et al. 2006), V0332+53
(Takeshima et al. 1994, Qu et al. 2005) and 4U 0115+63 (Soong
\& Swank 1989) while the Persistent HMXB pulsars with intermittent QPO
features have been detected are 4U 1907+09 (In't Zand et al. 1998, Mukerjee et al. 2001),
SMC X-1 (Angelini et al. 1991), Cen X-3 (Takeshima et al. 1991), LMC X-4 
(Moon \& Eikenberry 2001a, La Barbera et al. 2001) and X-Persei (Takeshima 1997). 
For most of the accretion powered pulsars the QPOs are a transient phenomenon. See Finger
(1998) for a review of the QPOs in transient X-ray pulsars and evolution of
the QPO feature along the X-ray outbursts. The QPO frequency in HMXB pulsars
detected so far have frequency in the range of 1 mHz to 400 mHz. This is the
highest frequency QPOs ever detected among the HMXB pulsars.
 
Several models have been proposed to explain the QPO generation mechanism
in accretion powered X-ray pulsars among which Keplerian Frequency Model
(KFM) and Beat Frequency Model (BFM) are used most frequently. Both KFM
(QPOs arise from the modulation of the X-rays by inhomogeneities in the
inner disk at the Keplerian frequency; van der Klis et al. 1987) and BFM
(oscillations occur at the beat frequency between orbital frequency of
matter in accretion disk at the Alfv\'{e}n radius and the stellar spin
frequency; Alpar \& Shaham 1985, Shibazaki \& Lamb 1987) are in good
agreement with X-ray pulsars EXO 2030+375 and A 0535+262. In X-ray pulsars
4U 0115+63, V 0332+52, Cen X-3, 4U 1626-67 and SMC X-1, the pulsar frequency
is higher than QPO frequency hence the KFM is not applicable in these
sources because if the Keplerian frequency at the magnetospheric boundary
is less than the spin frequency of the pulsar, propeller effect would inhibit
accretion. KFM and BFM also expect a positive correlation between QPO
centroid frequency and the X-ray luminosity of an X-ray pulsar, which is
not seen in some of the X-ray pulsars like V 0332+52 and GRO J1744-28.
Therefore, both KFM and BFM are not applicable in these sources. Recently,
Shirakawa \& Lai (2002) have shown that the low frequency QPOs in accreting
X-ray pulsars can also be due to a magnetically driven precession of warped
inner accretion disk. 

In XTE J0111.2--7317, both the KFM and BFM models are applicable and
as the spin frequency (0.032 Hz) is much smaller than the QPO frequency
(1.27 Hz), both these models would give similar value of the radius at
which the $\sim$1.3 Hz QPOs are produced. Assuming a neutron star mass
of 1.4$M_\odot$ the radius of the QPO production region is calculated
to be
$r_{QPO} = \left(\frac{GM_{NS}}{4\pi^2\nu^{2}_{k}}\right)^{1/3} = 1.4\times10^{8}$ cm.

The ASCA observations of this source during the outburst measured a flux
level of $3.6\times 10^{-10}$ ergs cm$^{-2}$s$^{-1}$ in the energy range
of 0.7-10.0 keV (Yokogawa et al. 2000) and BATSE/ASM on CGRO measured
similar pulsed flux in energy band of 20-50 keV. Assuming a pulse fraction
of about 50\%, the total X-ray flux of XTE J0111.2--7317 can be estimated
to be about 4 times of that found from the ASCA observations. As XTE J0111.2--7317
belongs to the Small Magellanic Cloud, the distance uncertainity is
relatively less compared to the Galactic X-ray binaries and we assume a source
distance of 65 kpc. Therefore, the total X-ray luminosity of the source
L$_{X}$ at 65 kpc is calculated to be about 7.3 $\times$10$^{38}$ erg s$^{-1}$.
The radius of the inner accretion disk around a magnetised neutron star with
mass of 1.4 M$\odot$ and a radius of 10 km, can be approximately expressed
in terms of its magnetic moment and X-ray luminosity as
$r_M = 3\times 10^8 L_{37}^{-2/7}\mu_{30}^{4/7}$ (Frank et al. 1992),
where, 
L$_{37}$ is the X-ray luminosity in the units of 10$^{37}$ erg and
$\mu_{30}$ is the magnetic moment in units of $10^{30}$ cm$^3$Gauss. 

	Assuming that the QPOs are produced at the inner accretion disk,
i.e. equating $r_{QPO}$ with $r_{M}$, the magnetic moment of the neutron
star is calculated to be 2.2$\times$10$^{30}$ Gauss cm$^3$, which for a 
neutron star radius of 10 km, is equivalent to a magnetic field strength 
in the range of 2.2 to 4.4 $\times$ 10$^{12}$ Gauss depending on the
magnetic latitude. The magnetic field strength in this 
pulsar is quite comparable to most other HMXB pulsars. In the absence of a 
cyclotron absorption feature (Coe et al. 1998) detected in the X-ray spectrum 
of this source, the QPO
frequency and the X-ray luminosity provides us with the only way to estimate
the magnetic field strength in this source. For neutron star magnetic field
strength of $\sim$ 10$^{12}$ Gauss, the energy of the cyclotron absorption
feature on the stellar surface is $\sim$ 11.6 keV, thus for this pulsar an
absorption feature is expected in the energy range 25-50 keV. 
More sensitive spectroscopic observations in the hard X-ray band during
future outbursts of this source will be useful to detect possible 
spectral feature due to a cyclotron absorption.

\section{Conclusions}

We have discovered X-ray Quasi-periodic oscillations at 1.27 Hz
during an outburst of
the transient high mass binary X-ray pulsar XTE J0111.2-7317. This is the
highest frequency QPO feature in this class of objects. We have ruled out
the possibility that the QPO feature is associated with the nearby bright
X-ray pulsar SMC X-1. Using the X-ray luminosity and the measured QPO
frequency and applying models in which the QPOs are produced because of
motion of inhomogeneous matter in the inner accretion disk we have estimated
the magnetic field strength of the neutron star, which is quite comparable to 
other pulsars of this class.

\section*{Acknowledgements} 
We thank an anonymous referee for suggestions that helped us to improve
the paper.

\clearpage
\begin{figure}
\centering
\includegraphics[height=6.5in, angle=-90]{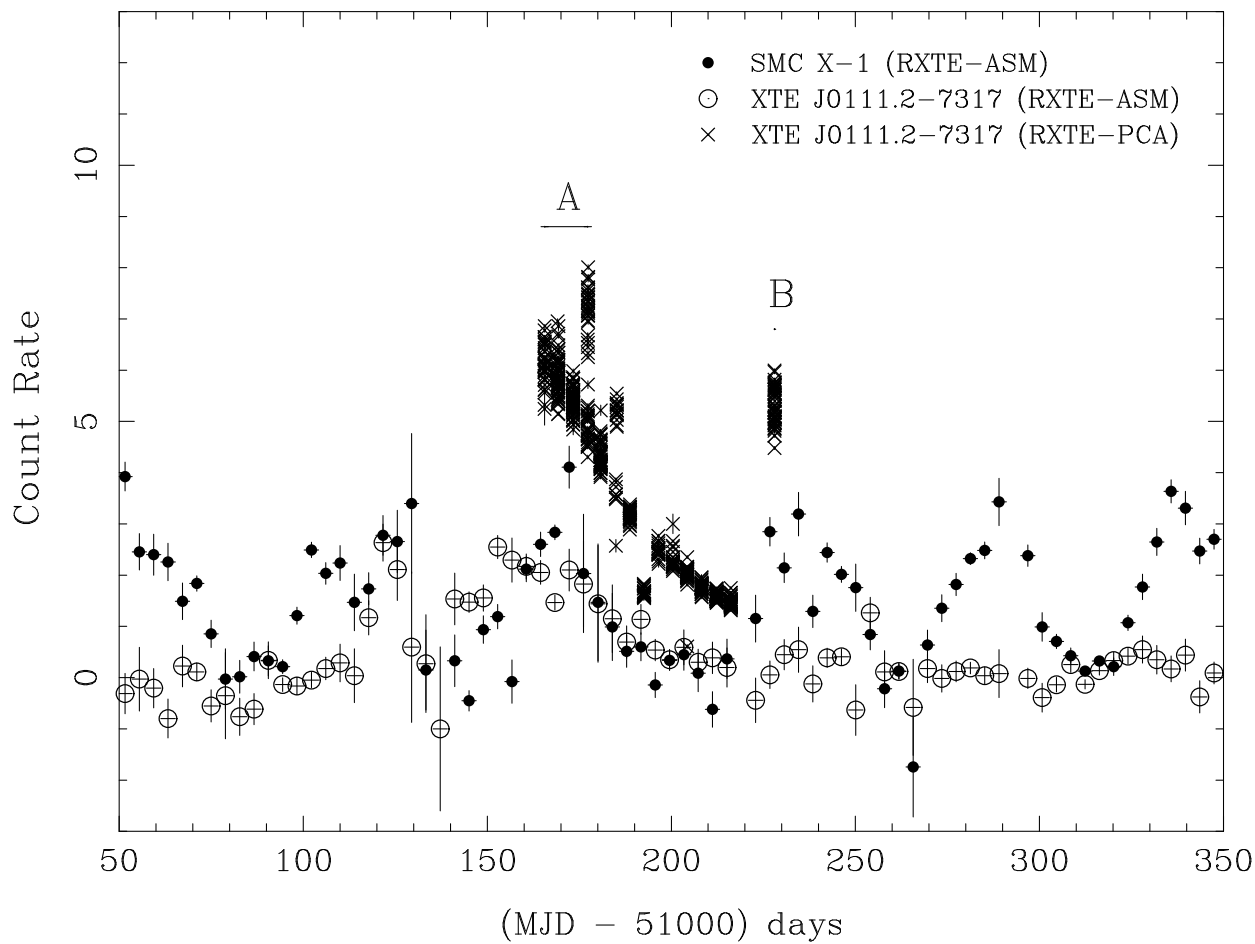}
\caption{RXTE-ASM Lightcurve of XTE J0111.2--7317 and SMC X-1 over a period
of 300 days are shown here along with rescaled RXTE-PCA lightcurve from
observations made towards XTE J0111.2--7317. Two time ranges MJD 51165.56 to
51177.32 and MJD 51228.00 to 51228.09 marked with 'A' and 'B' correspond to
the time when the two sources XTE J0111.2--7317 and XMC X-1 were bright
respectively. The two power density spectra for the segments 'A' and 'B'
are shown in Figure 2.}
\end{figure}

\begin{figure}
\centering
\includegraphics[height=6.5in, angle=-90]{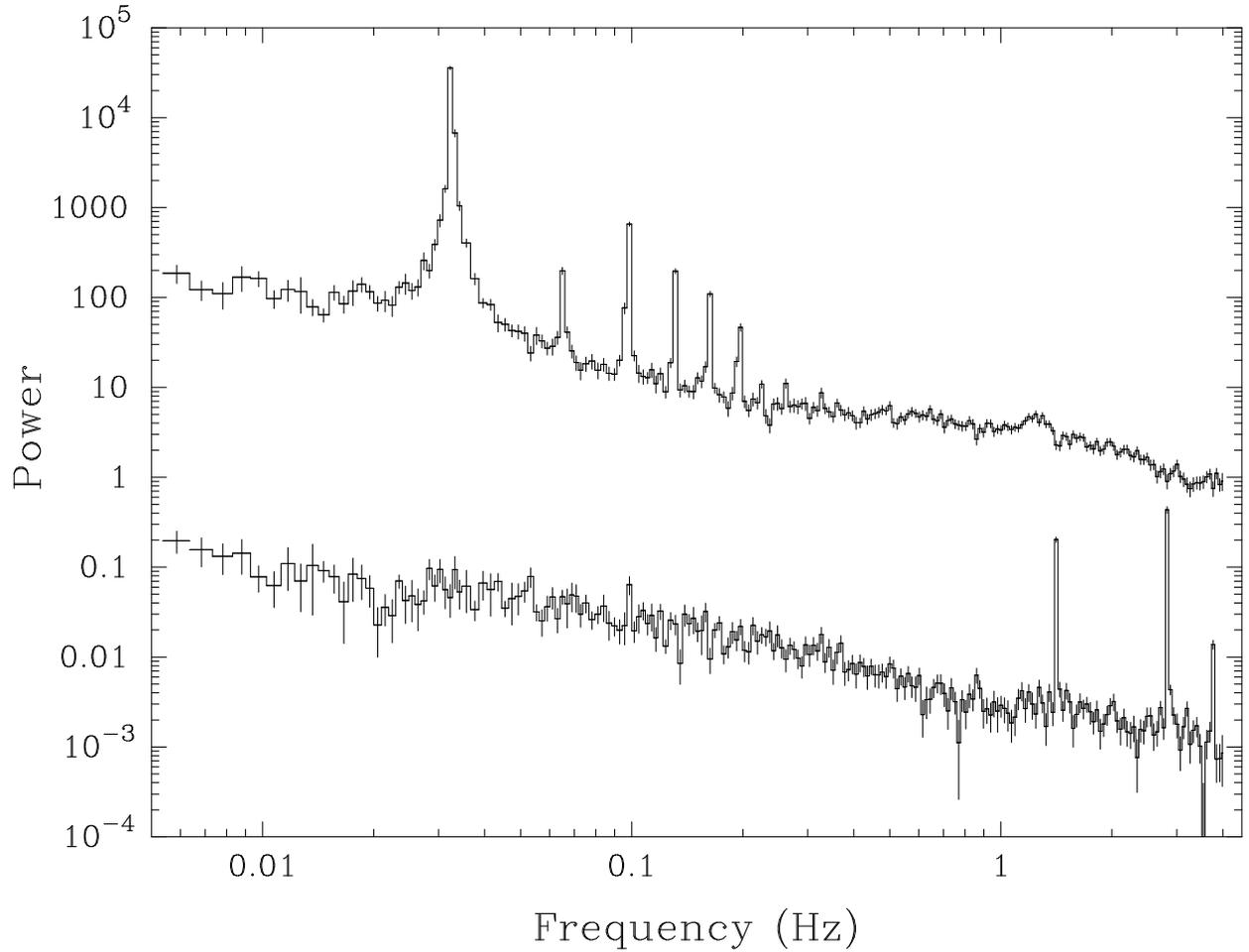}
\caption{Power density spectra generated from the lightcurves obtained
from RXTE-PCA observations made towards XTE J0111.2--7317 are shown here.
The top and bottom spectra are for the time ranges 'A' and 'B' respectively shown in
Figure 1. The top figure has been multiplied by a factor of 500 for
the sake of clarity.}
\end{figure}
\begin{figure}
\centering
\includegraphics[height=6.5in, angle=-90]{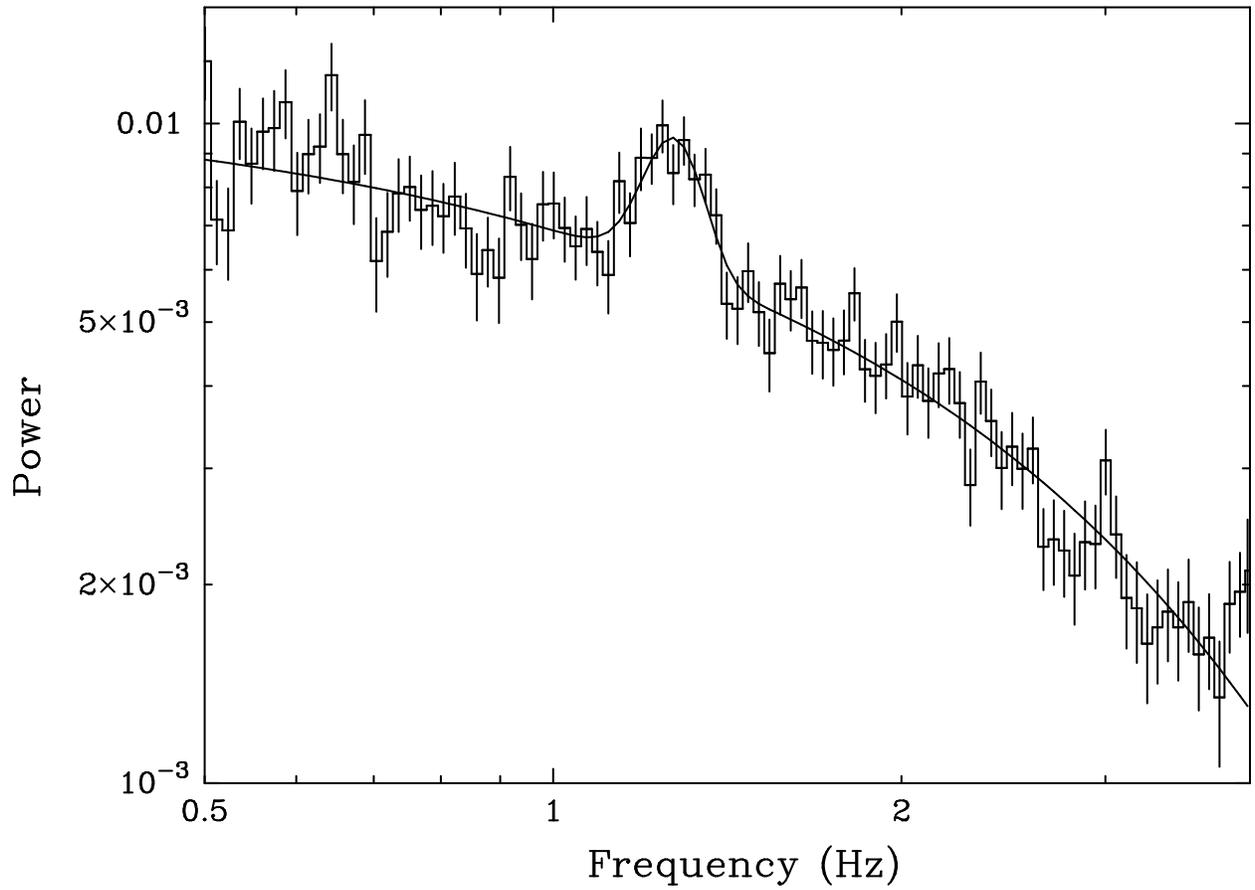}
\caption{Power density spectrum of XTE J0111.2--7317 generated from the lightcurve over the 
entire energy band of the PCA. The line represents the best fitted model for the contnuum and 
a Gaussian centered at the QPO frequency.} 
\end{figure}

\begin{figure}
\centering
\includegraphics[height=6.5in, angle=-90]{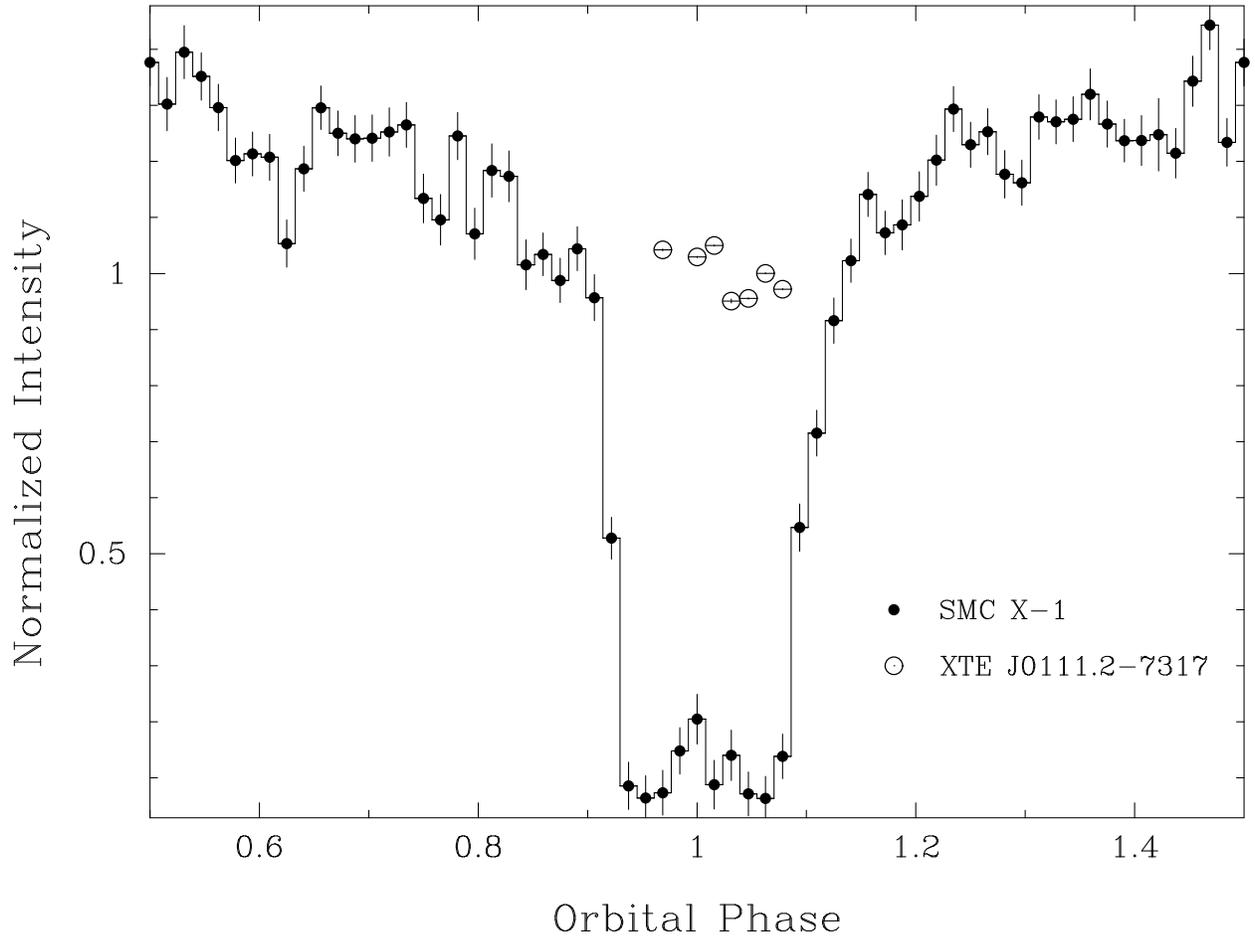}
\caption{RXTE-ASM light curve of SMC X-1 folded with its orbital period along with RXTE-PCA 
light curve of XTE J0111.2--7317 during segment A of Figure 1.}
\end{figure}

\begin{figure}
\centering
\includegraphics[height=6.5in, angle=-90]{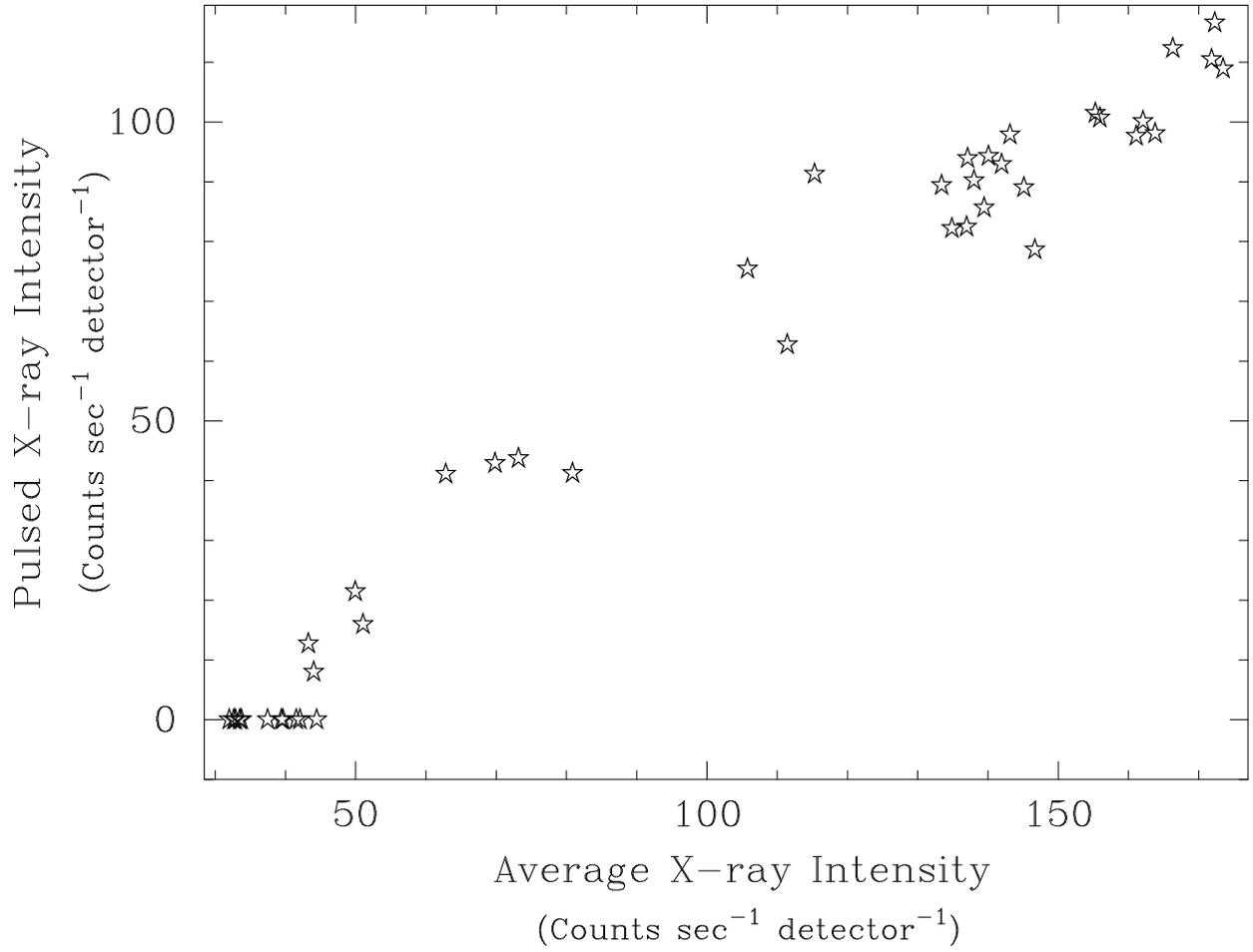}
\caption{Relation between average count rate and pulsed count rate for SMC X-1
is shown here. The formal correlation coefficient is determined to be 0.97.}
\end{figure}

\end{document}